\documentclass{iopart}

\usepackage{epsfig}

\newcommand{\be}{\begin{equation}}
\newcommand{\ee}{\end{equation}}
\newcommand{\bea}{\begin{eqnarray}}
\newcommand{\eea}{\end{eqnarray}}
\newcommand{\non}{\nonumber}

\begin{document}

\title{A new approach to non-commutative inflation}

\author{Massimiliano Rinaldi} 

\address{D\'epartment de Physique Th\'eorique, Universit\'e de Gen\`eve, \\ 24 quai E.\ Ansermet  CH--1211 Gen\`eve 4, Switzerland.}  
\ead{massimiliano.rinaldi@unige.ch}

\begin{abstract} We propose an inflationary scenario inspired by a recent formulation, in terms of coherent states, of  non-commutative quantum field theory. We consider the semiclassical Einstein equations, and we exploit the ultraviolet finiteness of the non-commutative propagator to construct the expectation value of the energy momentum tensor associated to a generic scalar field. It turns out that the latter is always finite and dominated by an effective cosmological constant. By combining this general feature with the intrinsic fuzziness of spacetime, we show that  non-commutativity governs the energy density of the early Universe in such a way that the strong energy condition is violated. Thus, there might be a bounce and a subsequent inflationary phase, which does not need any \emph{ad hoc} scalar field.
\end{abstract}


\section{Introduction}

\noindent The standard theory of big bang has achieved many successes, from explaining the observed thermal cosmic microwave background (CMB), to the expansion of the Universe and the abundance of light elements. However, it is an incomplete theory as it does not explain why the Universe is almost flat and so homogeneous on large scales. The inflationary paradigm was originally introduced to answer successfully these questions  \cite{Infl}. In addition, inflationary theories predicted the tiny fluctuations in the CMB, confirmed later on by observations  with increasing accuracy \cite{Komatsu}.

By definition, inflation consists in a period of accelerated expansion of the Universe, during which gravity acts as a repulsive force. In the simplest inflationary model, this effect is achieved if  the dynamics is dominated by a minimally coupled scalar field, called inflaton. With this hypothesis, the strong energy condition is violated, and the energy density behaves as a weakly time-dependent and positive cosmological constant. At the end of inflation, the inflaton decays into radiation and matter according to various reheating mechanisms \cite{books}. 

In this paper we investigate the possibility that the inflationary phase is not driven by a classical scalar field, but it rather  originates from a non-commutative structure of spacetime, which governs the dynamics through quantum effects encoded in the expectation value of the stress tensor. The idea that non-commutativity can drive the dynamics of the early Universe through its influence on the matter fields is not new, and it was utilized  in many important works, see e.g. \cite{BraMag}. 

According to the semiclassical picture, the Einstein equations are given by 
\bea\label{semic}
R_{\mu\nu}-{1\over 2}Rg_{\mu\nu}=8\pi G \langle T_{\mu\nu}\rangle\ ,
\eea
where $\langle T_{\mu\nu}\rangle$ is the expectation value of the energy momentum tensor associated to the matter fields. In general, this quantity is divergent but it can be formally written by acting with a differential operator on the propagator, expressed as a function of two distinct points \cite{BirDav}. For example, for a minimally coupled scalar field of mass $m$, and in Euclidean spacetime, we can write
\bea\label{tmunu}\non
\langle T_{\mu\nu}(x,x')\rangle &=& {1\over 2}\left(g^{\alpha'}_{\,\,\,\mu} \nabla_{\alpha'}\nabla_\nu +g^{\alpha'}_{\,\,\,\nu}\nabla_{\mu}  \nabla_{\alpha'}  \right)G_E(x,x')+\\
&-&{1\over 2}g_{\mu\nu}\left(g^{\alpha' \beta}\nabla_{\alpha'}\nabla_{\beta}+m^2\right)G_E(x,x')\ ,
\eea 
where primed indices refer to $x'$, and $G_E(x,x')$ is the Euclidean propagator \cite{Chris}. In the coincident limit $x=x'$ the propagator diverges, and so it does $\langle T_{\mu\nu}\rangle$. Generally speaking, the singularity structure of the two-point function can always be cast in the Hadamard form \cite{full}
\bea\label{Hada}
G(x,x')={u(x,x')\over \sigma(x,x')}+v(x,x')\ln |\sigma(x,x')| +w(x,x')\ ,
\eea
where $\sigma(x,x')$ is half the square geodesic distance between $x$ and $x'$, and $u$, $v$, and $w$ are finite functions in the coincident limit \cite{BirDav}.   The usual renormalization procedure brings along the addition of higher-curvature terms in the gravitational action, and the Einstein equations become extremely complicated, see e.g. \cite{Davies} for the cosmological case. Moreover, the right-hand-side of Eq.\ (\ref{semic})  might contain also a stochastic contribution when the fluctuations of the stress tensor are large \cite{stoch}, but we will not address this issue here. 

This situation might change if we modify sensibly the  Hadamard form of the propagator. For example, one can introduce a minimal length $L_P$ via  path-integral duality as in \cite{pad}. Then, the Euclidean Feynman propagator acquires a regulating term in the Schwinger representation, and takes the form  $G(x,x')=(\sigma(x,x')+L_P^2)^{-1}$. Connections between this proposal and  string T-duality were investigated in \cite{PadSpal}.

Another way of introducing a minimal length is via non-commutativity of spacetime \cite{ncfund}. Implementations of non-commutativity in inflationary models  were initially proposed by Brandenberger et al.\  \cite{Bran}, and Lizzi et al.\  \cite{Lizzi}. In these works, the effects of the non-commutativity of spacetime is encoded in the Moyal product, which replaces the ordinary product between functions in the Lagrangian of the inflaton field.

In this paper, we consider an alternative approach to non-commutativity, based on the coordinate coherent states proposed in \cite{Sma}. In particular, we would like to study what happens to the inflationary phase, when the intrinsic fuzziness of spacetime is expressed not by the Moyal product but by the Gaussian smearing of the source term in the Einstein equations, which is a direct effect of coherent state non-commutativity. In the next section, we will briefly review the main properties of the coordinate coherent states, and we will calculate the stress tensor of a scalar field via semiclassical arguments.  In Sec.\ (\ref{infla}), we will show how the energy density is modified by non-commutative effects, and how this leads to a non-singular solution  to the Friedmann equations. In Sec.\ (\ref{cc}) we study the case when also momenta are non-commuting, and how this can alter the value of the cosmological constant significantly. In the final section, we will discuss further aspects of this model, and the necessary steps to take in order to verify the consistency with current observations.

\section{Coherent states and non-commutativity}
\label{Cohe}

The  proposal put forward in \cite{Sma} introduces two novelties, with respect to the usual non-commutative scenario. First, all coordinate operators (and not only a subset of) must satisfy the relation $
[\hat z^{\mu},\hat z^{\nu}]=i\Theta^{\mu\nu} $, where $\Theta^{\mu\nu}$ is a constant and antisymmetric tensor. In  four Euclidean dimensions, one can transform this tensor in a block diagonal form, such that $\Theta^{\mu\nu}={\rm diag} (\theta_1 \varepsilon^{ij}, \theta_2 \varepsilon^{ij})$, where $\varepsilon^{ij}$ is the two-dimensional Levi-Civita tensor. It turns out that, if $\theta_1=\theta_2$, the resulting field theory is covariant. The second novelty is that physical coordinates are commuting c-numbers, constructed as expectation values on coherent states. For example, on the Euclidean plane we have two coordinate operators, which satisfy the algebra $[\hat z_1,\hat z_2]=i\theta$. In complete analogy with the case of the harmonic oscillator, where the roles of the position and the momentum operator are now played by the two coordinate operators, one can construct the ladder operators
\bea
 \hat A=\hat z_1+i\hat z_2\ , \quad  \hat A^{\dagger}=\hat z_1-i\hat z_2\ ,
\eea
such that $[\hat A,\hat A^{\dagger}]=2\theta$. The coherent states $|\alpha\rangle$ are defined as the ones which satisfy the equation $\hat A |\alpha\rangle =\alpha |\alpha\rangle$. Finally, the physical coordinates are the \emph{commuting} $c$-numbers 
 \bea 
 x_1={\rm Re}(\alpha)=\langle \alpha | \hat z_1 |\alpha \rangle\ , \quad  x_2={\rm Im}(\alpha)=\langle \alpha | \hat z_2 |\alpha \rangle\ .
\eea
Thus, on the non-commutative plane, the vector $(x_1,x_2)$ represents the mean position of a point-particle. We note that coherent states are not orthogonal, and form an over-complete basis \cite{Glauber}. In fact, if $|\alpha\rangle$ and $|\beta\rangle$ are coherent states, we have that
\bea\label{ortho}
\langle \alpha |\beta\rangle=\exp\left(\alpha^*\beta-{1\over 2}|\alpha|^2-{1\over 2}|\beta|^2\right)\ .
\eea
From this, it follows that any operator $\hat T$ has the expectation value 
\bea\label{opEV}
\langle \alpha |\hat T|\beta\rangle=\exp\left(-{1\over 2}|\alpha|^2-{1\over 2}|\beta|^2\right)T(\alpha^*,\beta)\ ,
\eea
where the function $T (\alpha^*,\beta)$ is defined as the double sum
\bea
T(\alpha^*,\beta)=\displaystyle\sum_{n,m}{T_{nm}\over\sqrt{n!m!}}(\alpha^*)^n\beta^m\ ,
\eea
which is well defined and convergent, see \cite{Glauber} for the proof. In the case of the wave operator $\exp (ip_1\hat z_1+ip_2\hat z_2)$, where $p_1$ and $p_2$ are the ordinary and commuting components of the momentum, one can easily compute the expectation value by using the Baker-Campbell-Hausdorff formula, and the result is \cite{Sma} 
\bea
\langle \alpha |e^{ip_1\hat z_1+ip_2\hat z_2}|\alpha\rangle=\exp \left[ i\vec p\cdot \vec x-{\theta\over 4}\left(p_1^2+p_2^2\right)\right]\ .
\eea
The Gaussian damping is the key feature of this approach to non-commutativity. The same damping enters in the integral measure of the Fourier transform operator, and from this one can construct the quantized theory of a scalar field, see e.g.\ \cite{maxlast}. The above construction can be lifted to higher and even-dimensional spacetimes, and the propagator in a four-dimensional Euclidean spacetime for a scalar field reads
\bea
G_E(p)={e^{-p^2\theta/4}\over p^2+m^2}\ .
\eea
In coordinate space, the above propagator is manifestly finite in the ultraviolet regime. In the massless limit, it takes the form \cite{MaxUnruh}
\bea\label{propag}
G_E(x,x')={1-e^{-\sigma_E(x,x')/4\theta}\over 4\pi^2\sigma_E(x,x')}\ ,
\eea
where $\sigma_E$ is the Euclidean geodesic distance. In the light of Eq.\ (\ref{tmunu}), we therefore expect that also $\langle T_{\mu\nu}\rangle$  will be modified by the non-commutative structure of spacetime, and we aim to investigate what this implies in the inflationary context. 

For simplicity, we consider a minimally coupled massive scalar field, which can be interpreted as an effective matter field, or as the field generated by vacuum fluctuations, with the mass acting as an infrared regulator.
First, we need to find an expression for the Euclidean propagator in curved space, and the simplest way is to use the proper-time Schwinger representation  
\bea\label{GE}
G_E(x,x')=\int_0^{\infty}ds K(x,x';s)\ ,
\eea
where $K=\langle x |e^{-i\hat H s} |x'\rangle $ and $\hat H$ is the Klein-Gordon operator \cite{BirDav}. In \cite{SpalTrace} it is shown that the modified kernel can be written as  $\exp(\theta \opensquare_x) K_0(x,x';s)$, where $K_0$ represents the kernel of the usual theory. As a result, at coincident points we have
\bea\label{kernel}\non
K(x;s)={e^{-m^2s}\over 16\pi^2 (s+\theta)^2} \Big[1
+e^{[s\theta/(s+\theta)]\opensquare}\sum_{n=1}^{\infty} s^n a_n(x)\Big]\ ,\\
\eea
where the deWitt-Schwinger coefficients $a_n(x)$ are combination of the Riemann tensor and its derivatives up to order $2n$, see also \cite{NicoSp}. Because of the minimal distance $\sqrt{\theta}$, the integral (\ref{GE}) is no longer singular and can be easily evaluated. By expanding the result for small $m$, we find that
\bea\label{gexp}
16\pi^2 G_E(x,x)={a_0\over \theta}-F_1(\theta m^2)a_1(x)+\cdots\ ,
\eea
where the dots indicate terms proportional to $a_2$ and $\opensquare a_1$ and $F_1=1+\gamma+\ln(\theta m^2)+{\cal O}(\theta m^2)$.  As mentioned above, the mass term acts as an infrared regulator. In this limit, the propagator looks like the one discussed in the case of path-integral duality \cite{pad}.
The other important ingredient is that the Euclidean Green's function satisfies, in curved space, the modified equation 
\bea\label{green}
(\opensquare_x +m^2)G_E(x,y)=\,e^{\theta \opensquare_x}\left[{1\over\sqrt{g}}\delta^{(4)}(x,y)\right]\ ,
\eea
as a result of the modification of the Fourier transform operator, as mentioned above \cite{SpalTrace}. Since the Green's function is finite at coincident points, we can write the expectation value of the stress tensor (\ref{tmunu}) as
\bea\label{tmunuexp}
\langle T_{\mu\nu}\rangle= \nabla_\nu\nabla_\mu G_E(x)-{1\over 2}g_{\mu\nu}\lim_{y\rightarrow x}e^{\theta \opensquare_x}\left[\delta^{(4)}(x,y)\over \sqrt{g}\right].
\eea
One can show that, for small $(y-x)^2/\theta$, we have the general formula \cite{SpalTrace}
\bea\label{identity}
e^{\theta \opensquare_x}\left[\delta^{(4)}(x,y)\over \sqrt{g}\right]={1\over 16\pi^2\theta^2}\,e^{-(x-y)^2/4\theta}+\cdots\ 
\eea
where the dots stand for terms proportional to powers of $R$, the Ricci scalar. From Eq.\ (\ref{gexp}), we see that the term $\nabla_\mu\nabla_\nu G_E$ is of adiabatic order four and higher. Thus, we conclude that the leading term of the deWitt-Schwinger expansion of the Euclidean stress tensor for a scalar field has the form
\bea\label{tmunueff}
\langle T_{\mu\nu}\rangle \simeq {1\over 32\pi^2\theta^2}\,g_{\mu\nu}\ .
\eea
This result is general and can be extended to higher spin fields. The non-commutative structure of spacetime naturally regulates the divergent ultra-violet behaviour of the propagator, so the stress tensor for matter fields is UV finite, and the leading term assumes the form of an effective cosmological constant, which diverges in the $\theta\rightarrow 0$ limit. That the cosmological constant is proportional to $\theta^{2}$ is not a surprise, as this is often the case when one introduces a UV cut-off $\theta$ as, for example, in effective field theories \cite{efftheor}.

\section{Inflationary phase}
\label{infla}

The main lesson learnt from these considerations is that when a minimal length is introduced as in Eq.\ (\ref{kernel}), there is a smearing effect on the expectation value of operators acting on the field. As a result, the quantum stress tensor has a universal form to the leading term.

It has been argued that the smearing effect of the non-commutative geometry can be encoded by replacing the mass term $M$ of the Schwarzschild solution,  with a Gaussian distribution \cite{Nico} 
\bea\label{subs}
M\delta(r)\longrightarrow \rho(r)={M\over (4\pi\theta)^{3/2}}\,e^{-r^2/4\theta}\ ,
\eea
where $r$ is the radial coordinate. The Gaussian function is then interpreted as the energy density of a conserved perfect fluid, with radial and tangential pressures. As a result, the Schwarzschild solution is modified into a non-singular solution, which matches asymptotically the usual one. Very recently, this formula was somewhat justified in terms of the Voros product, which is a relative of the star-product, and leads to a field theory that is equivalent to the one displayed above \cite{Banerjee}, see also \cite{scholtz}.

In this section, we would like to find an analogous formula in the case of a Robertson-Walker spacetime. As we are not able to calculate exactly the expression on the right hand side of Eq.\ (\ref{tmunuexp}) in curved spacetime, we use a sort of loop expansion.
According to Eq.\ (\ref{opEV}), if $\hat\rho$ represents the energy density operator acting on coherent states, one finds that
\bea\label{enden}
\langle\hat\rho\rangle=\langle \alpha | \hat \rho(\hat z_1,\hat z_2) |\alpha\rangle=e^{-|\alpha|^2}\,\rho(\alpha,\alpha^*)\ ,
\eea
and, in the case when $\hat\rho=\rho_0=$ const, 
\bea\label{consten}
\langle\hat\rho\rangle=e^{-(x_1^2+x_2^2)/4\theta}\rho_0\ .
\eea
In our case, we consider the result (\ref{tmunueff}) as the ``tree level'' approximation, and the $00$-component of this expression gives the constant energy density $\rho_{0}$. To obtain the energy density at one loop, we promote this constant to the tree-level energy density operator and we let it act on coherent states as in Eq.\ (\ref{consten}). In doing so, we must recall the global symmetries imposed by a homogeneous and isotropic Robertson-Walker background, that allows only for a time-dependent energy density. As a result of these considerations, we argue that the expression of the energy density to first order has the form 
\bea
\langle\hat\rho\rangle= \rho_0\,e^{-(t-t_0)^2/4\theta}\ ,
\eea
where $t_0$ is the origin of the time axis and $\rho_{0}=(32\pi^{2}\theta^{2})^{-1}$. Had we imposed spherical symmetry instead, we would have found  Eq.\ (\ref{subs}), provided we identify the black hole mass as the energy density, i.e. the 00-component of Eq.\ (\ref{tmunueff}). We recall that we are treating the semiclassical version of the Einstein equations, in the sense that we restrict the quantum effects on the right-hand side of these, as in Eq.\ (\ref{semic}). The exact treatment of the problem would involve the calculation of the expectation value the Einstein tensor over coherent states.

\begin{figure}
\centering
\includegraphics[width=9cm]{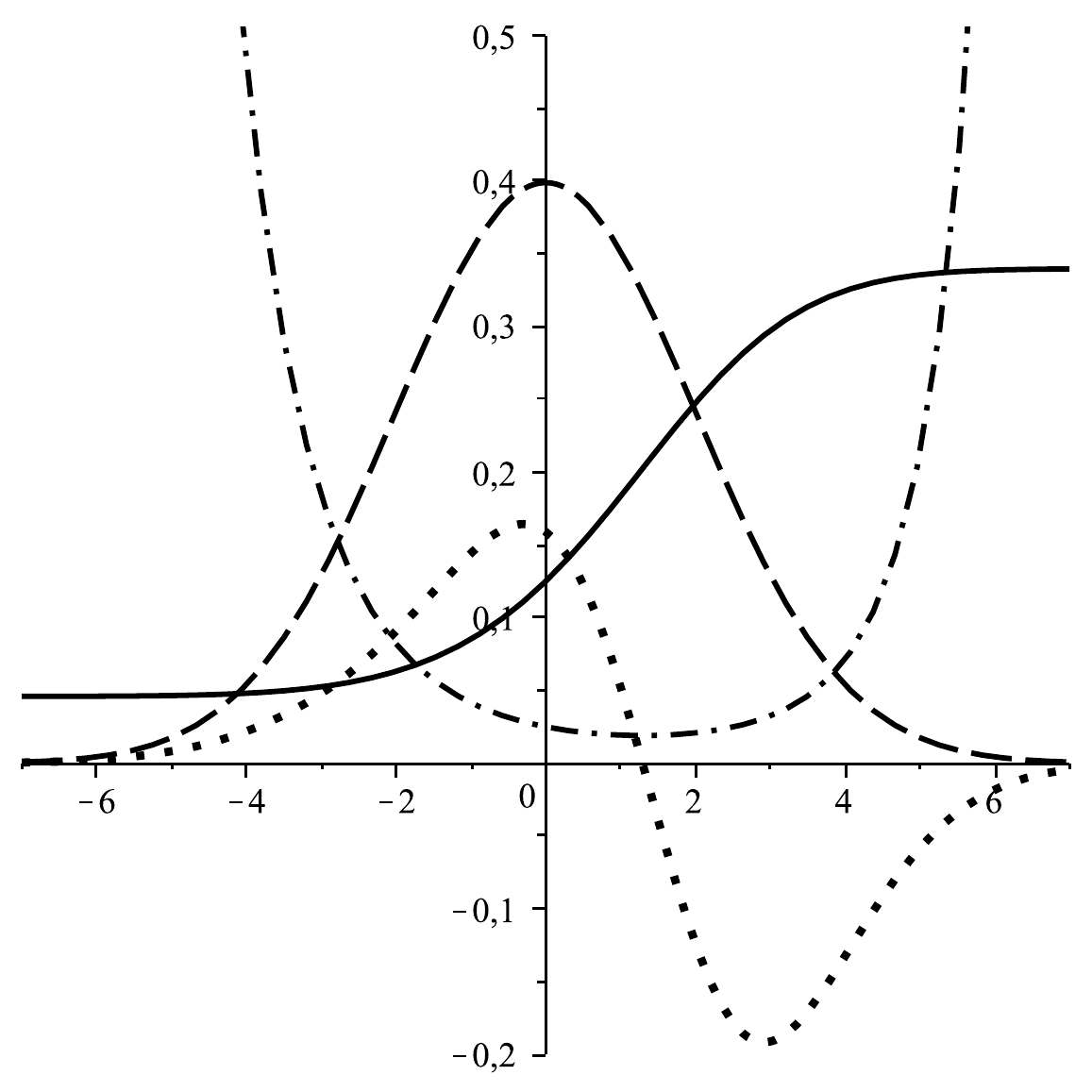} 
        \caption{Qualitative behaviour of $a$ (solid line), $\ddot a$ (dotted line), $H$ (dashed line), and $(aH)^{-1}$ (dot-dashed line) as functions of  time.}
	\label{FigA}
\end{figure}

If we consider the above equation as a good description of the quantum effects on the energy density, the Friedmann equation takes the form (we choose $t_0=0$)
\bea\label{NCFried}
\left(\dot a\over a\right)^2={8\pi G\over 3}\rho(t)\equiv H_0^2\, e^{-t^2/4\theta}\ ,
\eea
where $H=\dot{a}/a$ and $a$ is the scale factor of the metric $ds^2=-dt^2+a^2(t)\delta_{ij}dx^idx^j$. As the energy density is no longer singular, we can extend $t$ from $+\infty$ to $-\infty$. Then, the solution to Eq.\ (\ref{NCFried}) reads
\bea\label{scaleF}
a(t)=a_0\exp\left[{H_0\sqrt{2\pi\theta}}\,\, {\rm Erf}\left(t\over 2\sqrt{2\theta}\right)\right]\ ,
\eea
where
\bea
 {\rm Erf}(x')={2\over\sqrt{\pi}}\int_0^{x'}\,e^{-x^2}dx\ ,
\eea
and $a_0$ is an integration constant. It can be checked that $\ddot a$ is initially positive and then changes sign at the time when the comoving Hubble length $(aH)^{-1}$ reaches its minimum, while $H$ reaches a maximum value at $t=0$, see Fig.\ (\ref{FigA}).  In other words, the global evolution of the scale factor is very similar to the case of the pre-Big Bang scenario of string cosmology, as $\dot H(t)=-{t\over 4\theta H_0}H(t)$ \cite{Ven}.

The number of e-folds is $N=\ln (a_f/a_i)$, where the asymptotic scale factors are $a_{f,i}= a(t=\pm\infty)$. We find that $N=8\pi\sqrt{G\rho_0\theta\over 3}$ and, if $N\simeq 60$, this implies that $\sqrt{\theta}\simeq 0.014\, \ell_p$, where $\ell_p$ is the Planck length, and where we assumed that at $t=0$ the value of $\rho_0$ is given by the time-time component of Eq.\ (\ref{tmunueff}). Thus, the minimal length determines uniquely the number of e-folds in the specific case that matter is represented by a single scalar field. In more realistic models, there could be a large number $n$ of other fields that would change the result by a factor $\sqrt{n}$. In any case, the number of e-fold is basically determined by the minimal length of the theory.

The 00-component of the conservation equation for the energy-momentum tensor, yields $\dot\rho+3H(\rho+p)=0$, that can be used to find the effective pressure $p$. If we assume an equation of state of the form $p(t)=\omega(t)\rho(t)$, we find that
\bea
\omega\simeq -1+{t\over{6\theta H_0}}+{\cal O}\left({t^3\over\theta^2}\right)\ .
\eea
This leads to a dynamical crossing of the $\omega_0=-1$ value at the point where $H(t)$ reaches its maximum.

\section{Effective cosmological constant}
\label{cc}

If we interpret the result (\ref{tmunueff}) as a cosmological constant, we incur in the same problem as for standard field theory, namely that it is way too large when compared to the observed one. In fact, if we take $\theta$ to be of the order of the square of the Planck length, we would have the usual $10^{120}$ discrepancy problem  of the Standard Model \cite{weimbergrev}. One possibility is that $\theta$ is in fact only a ``bare'' parameter, that can be rescaled by an effective theory approach, along the lines of \cite{pad}. Another possibility is that this large quantity can be suppressed by properly defined boundary conditions \cite{michele}. There is however another intriguing way of dealing with this issue. The parameter $\theta$ can in fact be rescaled if we assume that also momenta do not commute. This hypothesis has been studied in several works (see e.g. \cite{NCmom} and references therein), and it is originally motivated by the interaction with a constant magnetic field $B$ of a particle moving on a non-commutative plane. In such a situation, the algebra of the operators is 
\bea
[\hat x_i,\hat x_j]=i\theta\varepsilon_{ij} \ ,\quad [\hat p_i,\hat p_j]=i\gamma\varepsilon_{ij} \ , \quad [\hat x_i,\hat p_j]=i\hbar\delta_{ij}\ ,
\eea
where $\varepsilon_{ij} $ is the two-dimensional Levi-Civita tensor, and $\gamma$ is a sort of ``minimal momentum'', which is proportional to $B$. This algebra can be reduced to the one with commuting momenta by rescaling the latter according to
\bea
\hat p_i\longrightarrow \hat \pi_i=\alpha \hat p_i+\beta \varepsilon_{ij} \hat x_j\ , \quad \alpha={1\over \sqrt{1-\theta\gamma}}\ , \\\non
 \beta={1\over\theta}(1-\alpha) \ ,\quad [\hat\pi_i,\hat\pi_j]=0\ .
\eea
We remark that, at the critical value $\gamma\theta=1$,  the linear transformation above becomes singular. This situation corresponds to a change of the symmetry group acting on the plane, from SU$(2)$ to SU$(1,1)$, as explained in \cite{NCmom}. In our context, the important fact is that when we compute the expectation value on coherent states of the rescaled wave operator $\exp(i\hat{\vec{\pi}}\cdot \hat {\vec x})$, we find that the damping factor is now multiplied by $\alpha$. It follows that the energy momentum tensor is modified according to
\bea
\langle T_{\mu\nu}\rangle\sim{1\over\theta^2\alpha^4}g_{\mu\nu}={(1-\gamma\theta)^2\over \theta^2}g_{\mu\nu}\ .
\eea
This shows that the effective cosmological constant can actually be much smaller than previously estimated, if the product $\gamma\theta$ is close to one.

%One could say that the expression above has a fine-tuning problem, as the observed value of the cosmological constant would be fixed by the arbitrary parameter $\gamma$. This is indeed true if $\gamma$ is a given constant. However, as we said before, this model is motivated by the interaction of a particle with an external magnetic field on a non-commutative plane. If we change the value of $B$, we also change $\gamma$. Therefore, we argue that, contrarily to $\theta$,  $\gamma$ is a time-dependent parameter, which depends on the cosmological expansion itself. In fact, one can think of $\gamma$ as the minimal momentum associated to the inverse of some maximal large distance scale. The most natural choice for this scale would be the (expanding) cosmological horizon size, which sets the largest possible causal distance. This is in line with the idea, proposed by Padmanabhan in \cite{Paddy2}, that the observed value of the cosmological constant depends on both the Planck length and the size of the causal horizon.  At the critical value $\gamma\theta=1$, the cosmological constant is zero and the symmetry  group on the non-commutative plane changes. This suggests that there might be a symmetry breaking mechanism, which is responsible for a non-zero, time-dependent cosmological constant. 

\section{Discussion}
\label{disc}

The inflationary phase ends when $\ddot a=0$, namely at the time $t_e$, implicitly determined by $t_e=4\theta H_0\,e^{-t^2_e/8\theta}$. The subsequent evolution of the Universe should be driven by a matter content other than the right hand side of Eq.\  (\ref{NCFried}). In most inflationary models, matter and radiation are generated by the decay of the inflaton. Alternatively, it is known that the expansion of the Universe itself can be a source of particle production \cite{ParProd}. As the expansion rate is initially of the order of $H_0\sim \theta^{-1/2}$, we can expect that the energy density of these particles will be roughly of the order of the Gibbons-Hawking temperature $T\simeq H_0/2\pi\sim \theta^{-1/2}$ \cite{Gibbons}. However, particle production could be tamed by the minimal length. In fact, in a recent study on the Unruh effect, based on the propagator  (\ref{propag}), it is shown that the thermal spectrum seen by an accelerated detector is strongly suppressed \cite{MaxUnruh}. This could be a hint towards excluding particle production as an efficient mechanism to produce matter in an expanding Universe, and further investigations are on their way.

There is another logical option that might render reheating not necessary. In fact, it seems plausible to assume that the Universe can collapse before $t=0$, because it is dominated by some form of (classical) matter. When the energy density grows and reaches planckian values, non-commutative effects take over and drive the Friedmann equations as explained above. In this sense, we can talk about a bouncing solution, in analogy with the pre-Big Bang scenario. In this scenario, the matter that dominates after the non-commutative phase can be a remnant of the primordial matter responsible for the crunch. 

Cosmological perturbations in the presence of a bounce have been studied since the early paper \cite{Starobinsky}. From these studies it turns out that quantum vacuum fluctuations, which exit the Hubble horizon during a matter-dominated contracting phase,  evolve to form a scale-invariant spectrum today  \cite{Bounce}. Eventually, a bouncing phase can leave specific non-gaussian signatures in the spectrum \cite{BrandBounce}. In our model, the final contracting phase is dominated by the non-commutative effects. On one hand,  one can argue that around the bounce at $t=0$, the Universe expand quasi-exponentially and therefore the background equations are indistinguishable from the ones already studied in the literature. It seems therefore plausible to claim that the spectrum of cosmological perturbation is scale invariant and that it might even show non-gaussianity. On the other hand, also perturbations feel the non-commutative structure of spacetime. Therefore, one should analyze the evolution of perturbations, taking in account that the propagator associated to graviton modes must satisfy an equation similar to (\ref{green}). Again, this could leave distinctive fingerprints in the CMB, and this exciting possibility will be the subject of a future work.

In this paper we described how the non-commutative structure of spacetime can act on the Einstein equations via semiclassical effects, and drive the Universe through an inflationary phase.  The semiclassical picture shows that the effect induced by non-commutativity on matter fields smears out the energy density, causing the violation of the strong energy condition, and, eventually, a smooth transition between a contracting and an expanding Universe. This phenomenon seems quite general, and it was already noticed in the context of non-commutative black holes. 

The inflationary scenario proposed in this work is based on two assumptions. The first is the algebra of the coordinate operators, that reminds the one for the coherent states of quantum mechanics. We mentioned that this model can be justified in terms of the Voros product, which belongs to the same family as the star-product \cite{Banerjee}. This means that both the star-product and the Voros product are equally valid candidates as \emph{the} non-commutative algebra. Therefore, assuming the Voros product is no more ad hoc than assuming the star-product as the algebra governing non-commutativity among coordinates.

The second, and perhaps the strongest, assumption made in this paper is that non-commutativity does not affect the left hand side of the Einstein equations as in the semiclassical approximation, where quantum effects are relegated to the source term only. This is common to many other non-commutative models and reflects our ignorance about a consistent way of quantizing the Einstein equations.  In our case, we assume that the quantum effects on the source arise from the fuzziness induced by quantum fluctuations at the scale $\theta$. A similar hypothesis  is adopted in \cite{BraMag} but we believe that our work has the advantage of going a little further, as it provides a mechanism for a regular bounce.

For these reasons, we believe that our model deserves further investigations. 
To begin with, it is crucial to determine the spectrum of the primordial fluctuations, an verify its flatness. Also, it is necessary to come up with a precise mechanism of reheating, or determine how matter eventually existing before the non-commutative phase evolves through it.  Finally, we should study more deeply the intriguing possibility that the effective cosmological constant arises from a symmetry-breaking at a more fundamental level of the theory, as mentioned in Sec.\ \ref{cc}.  We hope to report shortly and more accurately on these issues.

\ack

I wish to thank M.\ Arzano, C.\ B\oe hm, R.\ Durrer, P.\  Nicolini, and E.\ Spallucci for many valuable discussions. This work is supported by the Fond National Suisse.


\section*{References}

\begin{thebibliography}{99}


\bibitem{Infl} 
  V.~F.~Mukhanov and G.~V.~Chibisov,
  %``Quantum Fluctuation And Nonsingular Universe. (In Russian),''
  JETP Lett.\  {\bf 33} (1981) 532
  [Pisma Zh.\ Eksp.\ Teor.\ Fiz.\  {\bf 33} (1981) 549];
  %%CITATION = ZFPRA,33,549;%%
A.~A.~Starobinsky,
  %``Dynamics Of Phase Transition In The New Inflationary Universe Scenario And
  %Generation Of Perturbations,''
  Phys.\ Lett.\  B {\bf 117} (1982) 175.
  %%CITATION = PHLTA,B117,175;%%



\bibitem{Komatsu}
  E.~Komatsu {\it et al.}  [WMAP Collaboration],
  %``Five-Year Wilkinson Microwave Anisotropy Probe (WMAP\altaffilmark 1 )
  %Observations:Cosmological Interpretation,''
  Astrophys.\ J.\ Suppl.\  {\bf 180} (2009) 330.
  %%CITATION = APJSA,180,330;%%



\bibitem{books}
V.~Mukhanov,
  {\it Physical Foundations of Cosmology},
(Cambridge University Press, Cambridge, England, 2005);
R.~Durrer, 
{\it The Cosmic Microwave Background}
(Cambridge University  Press, Cambridge, England, 2008).

\bibitem{BraMag}
S.~Alexander, R.~Brandenberger and J.~Magueijo,
  %``Non-commutative inflation,''
  Phys.\ Rev.\  D {\bf 67} (2003) 081301.
  %%CITATION = PHRVA,D67,081301;%%

\bibitem{BirDav}
N.\ D.\ Birrell and P.\ C.\ W.\ Davies, {\it Quantum Fields in Curved Space}, (Cambridge University Press, London, 1982).

\bibitem{Chris}
S.~M.~Christensen,
  %``Vacuum Expectation Value Of The Stress Tensor In An Arbitrary Curved
  %Background: The Covariant Point Separation Method,''
  Phys.\ Rev.\  D {\bf 14} (1976) 2490.
  %%CITATION = PHRVA,D14,2490;%%



\bibitem{full}
R.~M.~Wald,
  %``The Back Reaction Effect In Particle Creation In Curved Space-Time,''
  Commun.\ Math.\ Phys.\  {\bf 54} (1977) 1;
  %%CITATION = CMPHA,54,1;%%
S.~A.~Fulling, M.~Sweeny and R.~M.~Wald,
  %``Singularity Structure Of The Two Point Function In Quantum Field Theory In
  %Curved Space-Time,''
  Commun.\ Math.\ Phys.\  {\bf 63} (1978) 257.
  %%CITATION = CMPHA,63,257;%%
  


  
\bibitem{Davies}
T.~S.~Bunch and P.~C.~W.~Davies,
  %``Nonconformal Renormalized Stress Tensors In Robertson-Walker Space-Times,''
  J.\ Phys.\ A  {\bf 11} (1978) 1315.
  %%CITATION = JPAGB,A11,1315;%%

\bibitem{stoch}
C.~I.~Kuo and L.~H.~Ford,
  %``Semiclassical gravity theory and quantum fluctuations,''
  Phys.\ Rev.\  D {\bf 47} (1993) 4510;
  %%CITATION = PHRVA,D47,4510;%%
E.~Calzetta and B.~L.~Hu,
  %``Noise and fluctuations in semiclassical gravity,''
  Phys.\ Rev.\  D {\bf 49} (1994) 6636.
  %%CITATION = PHRVA,D49,6636;%%



\bibitem{pad}
T.~Padmanabhan,
  %``Hypothesis of path integral duality. 1. Quantum gravitational corrections
  %to the propagator,''
  Phys.\ Rev.\  D {\bf 57} (1998) 6206.
  %%CITATION = PHRVA,D57,6206;%%

\bibitem{PadSpal}
 M.~Fontanini, E.~Spallucci and T.~Padmanabhan,
  %``Zero-point length from string fluctuations,''
  Phys.\ Lett.\  B {\bf 633} (2006) 627.
  %%CITATION = PHLTA,B633,627;%%

\bibitem{ncfund}
H.\ S.\ Snyders, Phys.\ Rev.\ {\bf 71}, 38 (1947);
A. Connes, {\it G\'eometrie non commutative}, (InterEditions, Paris, France, 1990). 


\bibitem{Bran} 
R.~Brandenberger and P.~M.~Ho,
  %``Noncommutative spacetime, stringy spacetime uncertainty principle, and
  %density fluctuations,''
  Phys.\ Rev.\  D {\bf 66} (2002) 023517.
  %%CITATION = 00346,12N1,10;%%

\bibitem{Lizzi} 
F.~Lizzi, G.~Mangano, G.~Miele and M.~Peloso,
  %``Cosmological perturbations and short distance physics from  noncommutative
  %geometry,''
  JHEP {\bf 0206} (2002) 049.
  %%CITATION = JHEPA,0206,049;%%
  


\bibitem{Sma}
A.\ Smailagic and E.~Spallucci,
  %``Lorentz invariance and unitarity in UV-finiteness of QFT on  noncommutative
  %spacetime,''
  J.\ Phys.\ A  {\bf 37} (2004) 1
  [Erratum-ibid.\  A {\bf 37} (2004) 7169];
  %%CITATION = JPAGB,A37,1;%%
  %``Feynman path integral on the noncommutative plane,''
  J.\ Phys.\ A  {\bf 36} (2003) L467;
  %%CITATION = JPAGB,A36,L467;%%
A.\ Smailagic and E.~Spallucci,
  %``UV divergence-free QFT on noncommutative plane,''
  J.\ Phys.\ A  {\bf 36} (2003) L517.
  %%CITATION = JPAGB,A36,L517;%%
  
  
\bibitem{Glauber}
R.~J.~Glauber,
  %``Coherent and incoherent states of the radiation field,''
  Phys.\ Rev.\  {\bf 131} (1963) 2766.
  %%CITATION = PHRVA,131,2766;%%
  
\bibitem{maxlast}
M.~Rinaldi,
  %``Particle production and transplanckian problem on the non-commutative
  %plane,''
  Mod.\ Phys.\ Lett.\  A {\bf 25} (2010) 2805;
  %%CITATION = MPLAE,A25,2805;%%
R.~Casadio, A.~Gruppuso, B.~Harms and O.~Micu,
  %``Boundaries and the Casimir effect in non-commutative space-time,''
  Phys.\ Rev.\  D {\bf 76} (2007) 025016.
  %%CITATION = PHRVA,D76,025016;%%

  
\bibitem{MaxUnruh}
P.~Nicolini and M.~Rinaldi,
  %``A minimal length versus the Unruh effect,''
  Phys.\ Lett.\  B {\bf 695} (2011) 303.
  %%CITATION = PHLTA,B695,303;%%


\bibitem{SpalTrace}
E.~Spallucci, A.~Smailagic and P.~Nicolini,
  %``Trace anomaly in quantum spacetime manifold,''
  Phys.\ Rev.\  D {\bf 73} (2006) 084004.
  %%CITATION = PHRVA,D73,084004;%%

\bibitem{NicoSp}
L.~Modesto and P.~Nicolini,
  %``Spectral dimension of a quantum universe,''
  Phys.\ Rev.\  D {\bf 81} (2010) 104040.
  %%CITATION = PHRVA,D81,104040;%%
  
\bibitem{efftheor}
A.~G.~Cohen, D.~B.~Kaplan and A.~E.~Nelson,
  %``Effective field theory, black holes, and the cosmological constant,''
  Phys.\ Rev.\ Lett.\  {\bf 82} (1999) 4971;
  %%CITATION = PRLTA,82,4971;%%
J.~F.~Donoghue,
  %``General Relativity As An Effective Field Theory: The Leading Quantum
  %Corrections,''
  Phys.\ Rev.\  D {\bf 50} (1994) 3874.
  %%CITATION = PHRVA,D50,3874;%%


\bibitem{Nico}
P.~Nicolini,
  %``A model of radiating black hole in noncommutative geometry,''
  J.\ Phys.\ A  {\bf 38}, L631 (2005);
  %%CITATION = JPAGB,A38,L631;%% 
P.~Nicolini, A.~Smailagic and E.~Spallucci,
%``Noncommutative geometry inspired Schwarzschild black hole,''
Phys.\ Lett.\  B {\bf 632} (2006) 547;
%%CITATION = PHLTA,B632,547;%%
P.~Nicolini,
  %``Noncommutative Black Holes, The Final Appeal To Quantum Gravity: A
  %Review,''
  Int.\ J.\ Mod.\ Phys.\  A {\bf 24} (2009) 1229;
  %%CITATION = IMPAE,A24,1229;%%
M.~Bleicher and P.~Nicolini,
 %``Large extra dimensions and small black holes at the LHC,''
  J.\ Phys.\ Conf.\ Ser.\  {\bf 237} (2010) 012008.
  %%CITATION = 00462,237,012008;%%

  
\bibitem{Banerjee}
  R.~Banerjee, S.~Gangopadhyay and S.~K.~Modak,
  %``Voros product, Noncommutative Schwarzschild Black Hole and Corrected Area
  %Law,''
  Phys.\ Lett.\  B {\bf 686} (2010) 181.
  %%CITATION = PHLTA,B686,181;%%

\bibitem{scholtz}
F.~G.~Scholtz, L.~Gouba, A.~Hafver and C.~M.~Rohwer,
  %``Formulation, Interpretation and Application of non-Commutative Quantum
  %Mechanics,''
  J.\ Phys.\ A  {\bf 42} (2009) 175303.
  %%CITATION = JPAGB,A42,175303;%%

  
\bibitem{weimbergrev}
  S.~Weinberg,
  %``The cosmological constant problem,''
  Rev.\ Mod.\ Phys.\  {\bf 61} (1989) 1.
  %%CITATION = RMPHA,61,1;%%

\bibitem{michele}
M.~Maggiore,
  %``Zero-point quantum fluctuations and the cosmological expansion,''
  arXiv:1004.1782 [astro-ph.CO].
  %%CITATION = ARXIV:1004.1782;%%

\bibitem{NCmom}

V.~P.~Nair and A.~P.~Polychronakos,
  %``Quantum mechanics on the noncommutative plane and sphere,''
  Phys.\ Lett.\  B {\bf 505} (2001) 267;
  %%CITATION = PHLTA,B505,267;%%
  S.~Bellucci, A.~Nersessian and C.~Sochichiu,
  %``Two phases of the non-commutative quantum mechanics,''
  Phys.\ Lett.\  B {\bf 522} (2001) 345.
  %%CITATION = PHLTA,B522,345;%%

  
\bibitem{Ven} 
M.~Gasperini and G.~Veneziano,
  %``The pre-big bang scenario in string cosmology,''
  Phys.\ Rept.\  {\bf 373} (2003) 1.
  %%CITATION = PRPLC,373,1;%%

  
\bibitem{ParProd}
L.~Parker,
  %``Quantized fields and particle creation in expanding universes. 1,''
  Phys.\ Rev.\  {\bf 183} (1969) 1057;
  %%CITATION = PHRVA,183,1057;%%
Y.~B.~Zeldovich,
  %``Particle production in cosmology,''
  Pisma Zh.\ Eksp.\ Teor.\ Fiz.\  {\bf 12} (1970) 443;
  %%CITATION = ZFPRA,12,443;%%
L.~H.~Ford,
  %``Gravitational Particle Creation and Inflation,''
  Phys.\ Rev.\  D {\bf 35} (1987) 2955.
  %%CITATION = PHRVA,D35,2955;%%




\bibitem{Gibbons}
  G.~W.~Gibbons and S.~W.~Hawking,
  %``Cosmological Event Horizons, Thermodynamics, And Particle Creation,''
  Phys.\ Rev.\  D {\bf 15} (1977) 2738.
  %%CITATION = PHRVA,D15,2738;%%

 \bibitem{Starobinsky}
  A.~A.~Starobinsky,
  %``Spectrum of relict gravitational radiation and the early state of the
  %universe,''
  JETP Lett.\  {\bf 30} (1979) 682
  [Pisma Zh.\ Eksp.\ Teor.\ Fiz.\  {\bf 30} (1979) 719].
  %%CITATION = ZFPRA,30,719;%%


  
 \bibitem{Bounce}
 F.~Finelli and R.~Brandenberger,
  %``On the generation of a scale-invariant spectrum of adiabatic  fluctuations
  %in cosmological models with a contracting phase,''
  Phys.\ Rev.\  D {\bf 65} (2002) 103522;
  %%CITATION = PHRVA,D65,103522;%%
D.~Wands,
  %``Duality invariance of cosmological perturbation spectra,''
  Phys.\ Rev.\  D {\bf 60} (1999) 023507;
  %%CITATION = PHRVA,D60,023507;%%
L.~E.~Allen and D.~Wands,
  %``Cosmological perturbations through a simple bounce,''
  Phys.\ Rev.\  D {\bf 70} (2004) 063515;
  %%CITATION = PHRVA,D70,063515;%%
P.~Peter and N.~Pinto-Neto,
  %``Cosmology without inflation,''
  Phys.\ Rev.\  D {\bf 78} (2008) 063506.
  %%CITATION = PHRVA,D78,063506;%%
  
  
\bibitem{BrandBounce}
Y.~F.~Cai, W.~Xue, R.~Brandenberger and X.~Zhang,
  %``Non-Gaussianity in a Matter Bounce,''
  JCAP {\bf 0905} (2009) 011.
  %%CITATION = JCAPA,0905,011;%%


\end{thebibliography}
\end{document}